\documentclass[preprint,nobibnotes,showpacs,showkeys]{revtex4}

\usepackage{graphicx}
\usepackage{amsmath}
\usepackage[latin1]{inputenc}
\usepackage{float}

\begin{document}

\author{A. N. de São José}
\affiliation{Departamento de Engenharia Elétrica,
Centro Federal de Educa{çã}o Tecnológica de Minas Gerais, 30510-000, Belo Horizonte, MG, Brazil}
\author{P. M. Dias}
\affiliation{Programa de Pós-Graduação em Modelagem Matemática e Computacional,
Centro Federal de Educa{çã}o Tecnológica de Minas Gerais, 30510-000, Belo Horizonte, MG, Brazil}
\author{A. R. Bosco de Magalh{ã}es \footnotemark[1]
\footnotetext[1]{Corresponding author. Tel.: +55 31 3319 6709; fax: +55 31 3319 6722.\newline
\emph{E-mail address}: arthur.magalhaes@pq.cnpq.br.}}
\author{J. G. Peixoto de Faria}
\affiliation{Programa de Pós-Graduação em Modelagem Matemática e Computacional,
Centro Federal de Educa{çã}o Tecnológica de Minas Gerais, 30510-000, Belo Horizonte, MG, Brazil}
\affiliation{Departamento de F{í}sica e Matemática, Centro Federal de Educa{çã}o Tecnoló%
gica de Minas Gerais, 30510-000, Belo Horizonte, MG, Brazil}

\title{Toward irreversibility with a finite bath of oscillators}

\begin{abstract}
We investigate the routes by which a bath composed of a finite number of oscillators 
at zero temperature approaches the induction of dissipation when it nears 
the usual limit of dense spectrum spread in an infinite interval. 
It is shown that, when this limit is taken, different distributions of 
environment frequencies can lead to the same irreversible evolution. However, 
when we move away from it, the dynamics departs from irreversibility in qualitatively different manners. 
\end{abstract}

\keywords{open quantum systems, quantum many-body models}
\pacs{03.65.Yz, 03.65.Fd}

\maketitle

\section{Introduction}

The description of macroscopic irreversible phenomena from microscopic 
reversible laws, in both quantum and classical domains, remains as one of the main challenges of the
statistical physics. 
In order to take into account dynamical processes
involving systems far from the equilibrium, like
thermalization, transport phenomena, 
and so on, one must resort to \textit{kinetic equations} \cite{spohn1980}.
In general, kinetic equations govern the effective dynamics of
a system of interest (a single particle, for example) 
interacting with a huge number of elementary external degrees of freedom of the environment where it is immersed.
A kinetic equation is obtained from the particular microscopic model 
describing the interaction between the system of interest and its external partners.
In classical domain, one usually assumes that the dynamics of the 
global system follows a Hamiltonian equation of motion whereas in quantum domain
it is governed by the Schr\"odinger equation.
Unfortunately, due to the enormous number of degrees of freedom to be considered, usually
the exact treatment of the microscopic
equations of motion is prohibitive and the use of approximations is necessary.
Thus, the correct understanding of the limits of validity of a given
kinetic equation depends on the physical conditions imposed to derive it.

As we remarked above, in the derivation of a quantum kinetic equation it is 
assumed the unitary evolution of the global system 
composed by the system of interest and the 
environmental degrees of freedom. If 
a quantum system has discrete eigenenergies, it is expected that
the evolved state will be found arbitrarily close of the initial state
\cite{bocchieri1957}. This result is the quantum version of the Poincaré 
recurrence theorem. Besides, it is expected that the
coarse-grained entropy of isolated, finite but large ensembles 
of quantum interacting particles exhibits a periodic or quasi-periodic
behavior \cite{percival1961}. In general, this last statement is valid for quantum systems 
with integrable classical analogues. So, for a quantum system obeying these
constraints there is a kind of quantum recurrence time which is so large
that the larger is such a system. Further, after the seminal work of 
Anderson \cite{anderson1958}, it is well-known the existence of systems where diffusion
is absent by virtue of the inhomogeneous distribution of the eigenenergies. 

The last comments illustrate the problem of the
extension of the $H$-theorem, established in classical statistical physics by
the Boltzmann's pioneer works, to the quantum domain (see, e.g., Ref. \cite{percival1961, 
neumann1929, goldstein2010}). The quantum $H$-theorem intends to set up the conditions
to the approach to thermal equilibrium by a macroscopic quantum system and consequently
irreversible processes involving such a system. In this work, we address the problem 
of the achievement of the irreversible effective dynamics of a 
system of interest, described by a quantum Markovian
master equation -- an important kind of quantum kinetic equation -- from the global unitary
(and, therefore, reversible) dynamics. Quantum master equations are largely
applied in several problems in physics, like quantum transport 
\cite{rau1996, forster1948, kenkre1974} and atom-photon interactions 
\cite{cohen1992, scully1997}. Further, since the seventies, the interest by them has
grown by virtue of the decoherence program, a theoretical framework devised
to give a suitable answer to the quantum measurement problem ascribing the
disappearance of macroscopic quantum correlations to the inevitable coupling
of a quantum system to its environment \cite{joos2003}.

Since the van Hove's work \cite{hove1955}, derivations 
of quantum master equations from first principles have 
shared a particular ingredient: the number of external degrees
of freedom is considered infinite, a limit that guarantees the irreversibility of the
dynamics of the system of interest. In general, one considers
the environment modeled by a thermal reservoir prepared in an initial state that characterizes 
thermal equilibrium. 
As discussed in Ref \cite{spohn1980}, due to the complexity of the
knotty action of the environment on the evolution of the system of interest, the dynamics of the
last can be considered a \textit{non-Markovian stochastic process}.
In fact, the evolved state of the global system is completely determined by its initial
state by virtue of the deterministic nature of the Schr\"odinger equation. Therefore,
the present state of the system of interest depends on the whole previous history of its evolution.
Under suitable limits, it is possible obtain a valid approximation
where the memory effects vanish. Several methods are employed to obtain
a quantum master equation from the particular microscopic model and we cite
the projection method developed by Nakagima \cite{nakagima1958}, Zwanzig \cite{zwanzig1964}, and
Mori \cite{mori1965}, the cummulant expansion \cite{kampen1974}, 
and the Feynman-Vernon influence functional
\cite{feynman1963, caldeira1987}. 

The weak coupling limit seems to be ubiquitous in 
the derivation of a quantum Markovian master equation from a microscopic model
and this approximation implies to consider the global state uncorrelated during all time.
However, there is not a uniform way to implement the necessary approximations in 
order to obtain a quantum master equation. In other words, the physical conditions
that give support to the approximations differ from one particular derivation for another.
This fact leads some authors to investigate the conditions that a proper Markovian quantum master 
equation could satisfy. These conditions concern, for example, the preservation
of the Robertson-Schr\"odinger uncertainty relations, the preservation of the
positivity, the existence of a Gibbsian stationary state, and the condition of detailed balance
\cite{dekker1984, alicki1989, faria1998}. Despite of this, Markovian quantum master equations
are successfully applied in the description of several phenomena and experiments, especially in
quantum optics \cite{cohen1992,scully1997,carmichael1999}. The 
dynamics of a damped mode of the electromagnetic field,
the decay of an excited atom by the emission of photons, the process
of parametric amplification are few examples of processes and phenomena
in quantum optics described by Markovian master equations. 
For a dissipative single mode of the electromagnetic field,
the system of interest is modeled by a harmonic oscillator, in the
case of an excited atom decaying to the ground state by
the emission of one photon, the system of interest is
modeled by a two-level system. In both cases, the environment is
modeled by a bath of harmonic oscillators in thermal equilibrium. 
As a result of the Markovian approximation, the models of damped
mode and unstable atom predict
\textit{exponential decay} of the excitation of the system of interest,
which is closely related with the Weisskopf-Wigner theory of
radiation decay \cite{weisskopf1930}. This behavior of the probability of survival of the excitation
as function of time yields a Lorentz-type spectral line that was introduced 
by Breit and Wigner \cite{breit1936} in order to describe the broadening
of resonance lines in particle scattering. Despite of fitting properly the experimental
data, the exponential decay is a byproduct of the Markovian 
approximation, \textit{i.e.}, the exponential law only approximately describes
the temporal dependence of the probability of survival of the excitation in
the system of interest. In fact, one expects that the decay of an excitation in quantum mechanics
depends quadratically on the time for very short times and is governed by a power-law for long times
\cite{hellund1953, facchi1999}.

The remarks presented above naturally bring some questions to the fore. 
The van Hove's solution to the problem of irreversibility makes use of the
limit of infinite number of external degrees of freedom interacting with a
quantum system. However, despite the huge number of elementary 
quantum systems that form a macroscopic object, this number is finite.
Therefore, if the system of interest is linearly coupled to its external
partners, some kind of recurrence may happen even though the time necessary
for this is possibly greater than any relevant characteristic time involved. 
On the other hand, the region of recurrence could be accessible in
experiments involving mesoscopic systems.
In this work, we intend to gain some insight on the achievement of the
effective irreversible dynamics of a system of interest coupled to external degrees
of freedom in the quantum domain. In order to do this, we chose as a laboratory the model of a quantum 
harmonic oscillator linearly coupled to a \textit{finite} bath of harmonic oscillators in the
rotating wave approximation (RWA). These external oscillators are prepared in the vacuum state,
simulating a bath at null temperature, and the main oscillator is prepared in a coherent state.
For this case, we can identify the conditions on the number of external
oscillators, the distribution of their frequencies along a finite bandwidth
around the frequency of the main oscillator, and the strength of the
coupling constants in order to obtain a behavior that resembles dissipation
within an intermediate range of time. All results were obtained numerically
in the same sense that was done in Ref. \cite{rivas2010} with some important 
differences. Here, we are interested in studying the conditions that
lead to an irreversible dynamics; there, the authors study the 
limits of validity of the assumptions
generally employed in the derivation of a quantum master equation in the
Markovian limit. The Markovian limit is also discussed here by the fitting
of the exponential decay of the energy stored in the main oscillator.

This work is organized as follows: in the next section, we present the model
and determine the dynamics of the relevant quantities. In Section III, we study
how the relevant parameters, like strength of the coupling constants, 
number of external oscillators and their distribution along a finite bandwidth
must be adjusted in order to produce an effective dynamics that resembles 
irreversibility. Section IV is reserved to conclusions and final comments.

\section{Model and dynamics}

The model is composed of an oscillator $O_{1}$ linearly coupled to a set of
oscillators $O_{j}$ ($j=2$ to $N$). The oscillator $O_{1}$ corresponds to
the system of interest and the other ones play the role of the environment.
The Hamiltonian is 
\begin{equation}
\hat{H}=\hat{H}_{0}+\hat{H}_{int},\qquad \hat{H}_{0}=\underset{k=1}{\overset{%
N}{\sum }}\hbar v_{k}\hat{a}_{k}^{\dagger }\hat{a}_{k},\qquad H_{int}=%
\underset{k=2}{\overset{N}{\sum }}\hbar g_{k}\left( \hat{a}_{1}^{\dagger }%
\hat{a}_{k}+\hat{a}_{k}^{\dagger }\hat{a}_{1}\right) ,
\label{Hcampocampo}
\end{equation}
where $\hat{a}_{k}^{\dagger }$ and $\hat{a}_{k}$ are creation and
annihilation bosonic operators, respectively, and $v_{k}$ and $g_{k}$ are
real coefficients. By displaying this Hamiltonian in the matrix form 
\begin{equation}
\hat{H}=\hbar \left( 
\begin{array}{cccc}
\hat{a}_{1}^{\dagger } & \hat{a}_{2}^{\dagger } & \cdots & \hat{a}%
_{N}^{\dagger }
\end{array}
\right) \left( 
\begin{array}{cccc}
v_{1} & g_{2} & \cdots & g_{N} \\ 
g_{2} & v_{2} & 0 & 0 \\ 
\vdots & 0 & \ddots & \vdots \\ 
g_{N} & 0 & \cdots & v_{N}
\end{array}
\right) \left( 
\begin{array}{c}
\hat{a}_{1} \\ 
\hat{a}_{2} \\ 
\vdots \\ 
\hat{a}_{N}
\end{array}
\right) ,
\end{equation}
and defining 
\begin{equation}
\mathbf{M=}\left( 
\begin{array}{cccc}
M_{1,1} & M_{1,2} & \cdots & M_{1,N} \\ 
M_{2,1} & M_{2,2} & \cdots & M_{2,N} \\ 
\vdots & \vdots & \ddots & \vdots \\ 
M_{N,1} & M_{N,2} & \cdots & M_{N,N}
\end{array}
\right)
\end{equation}
as a real orthogonal matrix satisfying 
\begin{equation}
\mathbf{M}^{T}\left( 
\begin{array}{cccc}
v_{1} & g_{2} & \cdots & g_{N} \\ 
g_{2} & v_{2} & 0 & 0 \\ 
\vdots & 0 & \ddots & \vdots \\ 
g_{N} & 0 & \cdots & v_{N}
\end{array}
\right) \mathbf{M}=\left( 
\begin{array}{cccc}
\omega _{1} & 0 & \cdots & 0 \\ 
0 & \omega _{2} & 0 & 0 \\ 
\vdots & 0 & \ddots & \vdots \\ 
0 & 0 & \cdots & \omega _{N}
\end{array}
\right) ,
\end{equation}
it becomes easy to see that the normal modes operators are obtained through
the variables transformation 
\begin{equation}
\left( 
\begin{array}{c}
\hat{A}_{1} \\ 
\hat{A}_{2} \\ 
\vdots \\ 
\hat{A}_{N}
\end{array}
\right) =\mathbf{M}^{T}\left( 
\begin{array}{c}
\hat{a}_{1} \\ 
\hat{a}_{2} \\ 
\vdots \\ 
\hat{a}_{N}
\end{array}
\right) ,  \label{A=Ma}
\end{equation}
since Hamiltonian (\ref{Hcampocampo}) can be written as 
\begin{equation}
H=\hbar \left( 
\begin{array}{cccc}
\hat{A}_{1}^{\dagger } & \hat{A}_{2}^{\dagger } & \cdots & \hat{A}%
_{N}^{\dagger }
\end{array}
\right) \left( 
\begin{array}{cccc}
\omega _{1} & 0 & \cdots & 0 \\ 
0 & \omega _{2} & 0 & 0 \\ 
\vdots & 0 & \ddots & \vdots \\ 
0 & 0 & \cdots & \omega _{N}
\end{array}
\right) \left( 
\begin{array}{c}
\hat{A}_{1} \\ 
\hat{A}_{2} \\ 
\vdots \\ 
\hat{A}_{N}
\end{array}
\right) =\hbar \underset{k=1}{\overset{N}{\sum }}\omega _{k}^{\dagger }\hat{A%
}_{k}^{\dagger }\hat{A}_{k}.
\end{equation}
The orthogonality of $\mathbf{M}$ can be used to show that the\ operators $%
\hat{A}_{k}^{\dagger }$ and $\hat{A}_{k}$ obey the usual commutation
relations for bosons. The vacuum of the normal modes is the same as that of
the original modes, what is easily proved by noting that the annihilation operators of the normal modes
are linear combinations of the annihilation operators of the original modes.

Let us assume that the initial state of the whole system is given by 
\begin{equation}
\left| \psi \left( 0\right) \right\rangle =\left| \alpha _{1},\alpha
_{2},...,\alpha _{N}\right\rangle ,  \label{inistate1}
\end{equation}
where the vector state $\left| \alpha _{1},\alpha _{2},...,\alpha
_{N}\right\rangle $ indicates that oscillator $O_{k}$ is in the coherent
state with amplitude $\alpha _{k}$. Using Eq. (\ref{A=Ma}), we see that the
coherent states related to the original operators $\hat{a}_{k}$ are also
coherent states associated to the normal modes operators $\hat{A}_{k}$: 
\begin{eqnarray}
\left| \alpha _{1},\alpha _{2},...,\alpha _{N}\right\rangle &=&\exp \left( 
\underset{k=1}{\overset{N}{\sum }}\left( \alpha _{k}\hat{a}_{k}^{\dagger
}-\alpha _{k}^{\ast }\hat{a}_{k}\right) \right) \left| 0,0,...,0\right\rangle
\notag \\
&=&\exp \left( \underset{k=1}{\overset{N}{\sum }}\left( \Lambda _{k}\hat{A}%
_{k}^{\dagger }-\Lambda _{k}^{\ast }\hat{A}_{k}\right) \right) \left|
0,0,...,0\right\rangle  \notag \\
&=&\left| \Lambda _{1},\Lambda _{2},...,\Lambda _{N}\right\rangle ,
\end{eqnarray}
where $\left| \Lambda _{1},\Lambda _{2},...,\Lambda _{N}\right\rangle $ is
a coherent state related to the normal modes with amplitudes 
\begin{equation}
\Lambda _{k}=\underset{j=1}{\overset{N}{\sum }}\alpha _{j}M_{jk}.
\end{equation}
Thus, 
\begin{eqnarray}
\left| \psi \left( t\right) \right\rangle &=&\left| \Lambda _{1}e^{-i\omega
_{1}t},\Lambda _{2}e^{-i\omega _{2}t},...,\Lambda _{N}e^{-i\omega
_{N}t}\right\rangle  \notag \\
&=&\exp \left( \underset{k=1}{\overset{N}{\sum }}\left( \Lambda
_{k}e^{-i\omega _{k}t}A_{k}^{\dagger }-\Lambda _{k}^{\ast }e^{i\omega
_{k}t}A_{k}\right) \right) \left| 0,0,...,0\right\rangle  \notag \\
&=&\exp \left( \underset{k=1}{\overset{N}{\sum }}\alpha _{k}\left( t\right)
a_{k}^{\dagger }-\alpha _{k}^{\ast }\left( t\right) a_{k}\right) \left|
0,0,...,0\right\rangle  \notag \\
&=&\left| \alpha _{1}\left( t\right) ,\alpha _{2}\left( t\right) ,...,\alpha
_{N}\left( t\right) \right\rangle ,
\end{eqnarray}
where 
\begin{equation}
\alpha _{k}\left( t\right) =\underset{j=1}{\overset{N}{\sum }}\underset{l=1}{%
\overset{N}{\sum }}\alpha _{j}\left( 0\right) M_{j,l}M_{k,l}e^{-i\omega
_{l}t}.
\end{equation}
Since all oscillators remain in coherent states, the environment does not
produce decoherence on the system. If the central oscillator starts in the
coherent state with $\alpha _{1}=1$ and the environment is in vacuum state,
the energy of oscillator $O_{1}$ evolves as 
\begin{equation}
E_{1}\left( t\right) =\left\langle \hbar v_{1}a_{1}^{\dagger
}a_{1}\right\rangle \left( t\right) =\hbar v_{1}\left| \alpha _{1}\left(
t\right) \right| ^{2}=\hbar v_{1}\left| \overset{N}{\underset{k=1}{\sum }}%
M_{1,k}^{2}e^{-i\omega _{k}t}\right| ^{2}.  \label{Ec}
\end{equation}
As can be shown by using results in Ref. \cite{magalhaes1}, expression (\ref
{Ec}) is also valid for the central oscillator starting in the Fock state with one 
excitation and the environment in vacuum. Unlike the coherent
state, for such initial condition the system loses purity.

\section{The way to irreversibility}

From now on, the system of units is chosen so as $\hbar v_{1}=1$, and
the values of all physical quantities below are given with respect to this system.
If the environment is composed of a set of oscillators with equally spaced
frequencies varying in the range from $1+p/2$ to $1-p/2$, and the coupling
constants are equal, decaying with $1/\sqrt{N-1}$ when the number of
environmental modes grows, we get, for even $N$, 
\begin{equation}
\mathbf{H}=\hbar \left( 
\begin{array}{cccc}
v_{1} & g_{2} & \cdots  & g_{N} \\ 
g_{2} & v_{2} & 0 & 0 \\ 
\vdots  & 0 & \ddots  & \vdots  \\ 
g_{N} & 0 & \cdots  & v_{N}
\end{array}
\right) =\left( 
\begin{array}{ccccccccc}
1 & \frac{g}{\sqrt{N-1}} & \frac{g}{\sqrt{N-1}} & \frac{g}{\sqrt{N-1}} & 
\frac{g}{\sqrt{N-1}} & \frac{g}{\sqrt{N-1}} & \cdots  & \frac{g}{\sqrt{N-1}}
& \frac{g}{\sqrt{N-1}} \\ 
\frac{g}{\sqrt{N-1}} & 1 & 0 & 0 & 0 & 0 & \cdots  & 0 & 0 \\ 
\frac{g}{\sqrt{N-1}} & 0 & 1+\mu  & 0 & 0 & 0 & \cdots  & 0 & 0 \\ 
\frac{g}{\sqrt{N-1}} & 0 & 0 & 1-\mu  & 0 & 0 & \cdots  & 0 & 0 \\ 
\frac{g}{\sqrt{N-1}} & 0 & 0 & 0 & 1+2\mu  & 0 & \cdots  & 0 & 0 \\ 
\frac{g}{\sqrt{N-1}} & 0 & 0 & 0 & 0 & 1-2\mu  & \cdots  & 0 & 0 \\ 
\vdots  & \vdots  & \vdots  & \vdots  & \vdots  & \vdots  & \ddots  & \vdots 
& \vdots  \\ 
\frac{g}{\sqrt{N-1}} & 0 & 0 & 0 & 0 & 0 & \cdots  & 1+\frac{p}{2} & 0 \\ 
\frac{g}{\sqrt{N-1}} & 0 & 0 & 0 & 0 & 0 & \cdots  & 0 & 1-\frac{p}{2}
\end{array}
\right) ,  \label{Hparticular1}
\end{equation}
where $\mu =p/\left( N-2\right) $. The form of $g_{k}$ is analogous to the
one assumed in master equation derivations in order to achieve finite
environmental effects when the thermodynamic limit $N\longrightarrow \infty $
is taken (see, for example, Ref. \cite{magalhaes2}).

\begin{figure}[th]
\vspace{-0.5cm} \hspace{0cm} 
\includegraphics[scale=0.5]{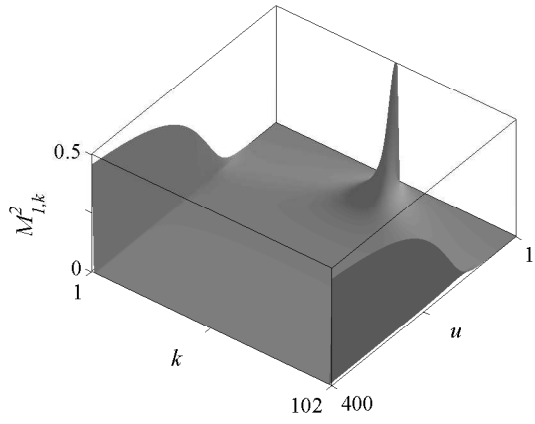}
\includegraphics[scale=0.5]{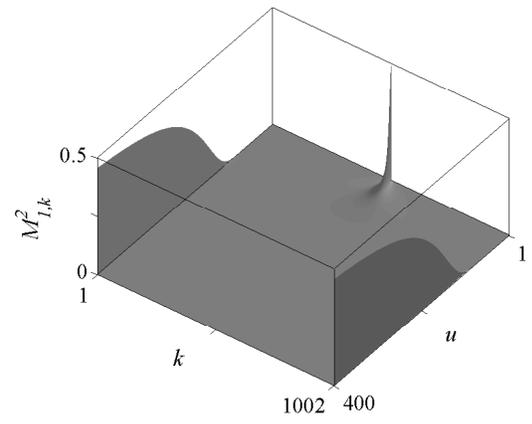} 

\textbf{(a)} \hspace{7cm} \textbf{(b)}

\vspace{-0.0cm}
\caption{Dependence of the set $\left\{ M_{1,k}^{2}\right\} $ as a function of $g=u/400$ 
($u$ ranging from $1$ to $400$)
for $\mathbf{H}$
given by Eq. (\ref{Hparticular1}) with $p=1$ and (a) $N=102$, (b) $N=1002$.}
\label{Mvar100}
\end{figure}

The dynamics of $E_{1}\left( t\right) $ depends on the distributions of $%
M_{1,k}^{2}$ and $\omega _{k}$, which, although usually calculated only
numerically, can be understood by considering the general situation as
intermediate between analytically tractable cases. In Figs. (\ref{Mvar100}),
we exemplify the change-over of the set $\left\{
M_{1,k}^{2}\right\} $.
For small $g/p$, the matrix in expression (\ref
{Hparticular1}) approximates a diagonal matrix with a two fold degenerate eigenvalue $1$,
leading to $M_{1,N/2}^{2}=M_{1,N/2+1}^{2}=1/2$ and $M_{1,k}^{2}=0$
otherwise. For large $g/p$, the set $\left\{ M_{1,k}^{2}\right\} $ becomes
like the one for $p=0$, where the only nonzero elements are $%
M_{1,1}^{2}=M_{1,N}^{2}=1/2$; this can be understood by noticing that if $%
\mathbf{M}$ diagonalizes $\mathbf{H}$ then it also diagonalizes $\mathbf{H}/g
$, whose diagonal elements approach each other when $g/p$ increases. When
the ratio $g/p$ grows from zero, $\left\{ M_{1,k}^{2}\right\} $ varies from
the former situation to the latest. 

\begin{figure}[th]
\vspace{-0.5cm} \hspace{0cm} 
\includegraphics[scale=0.5]{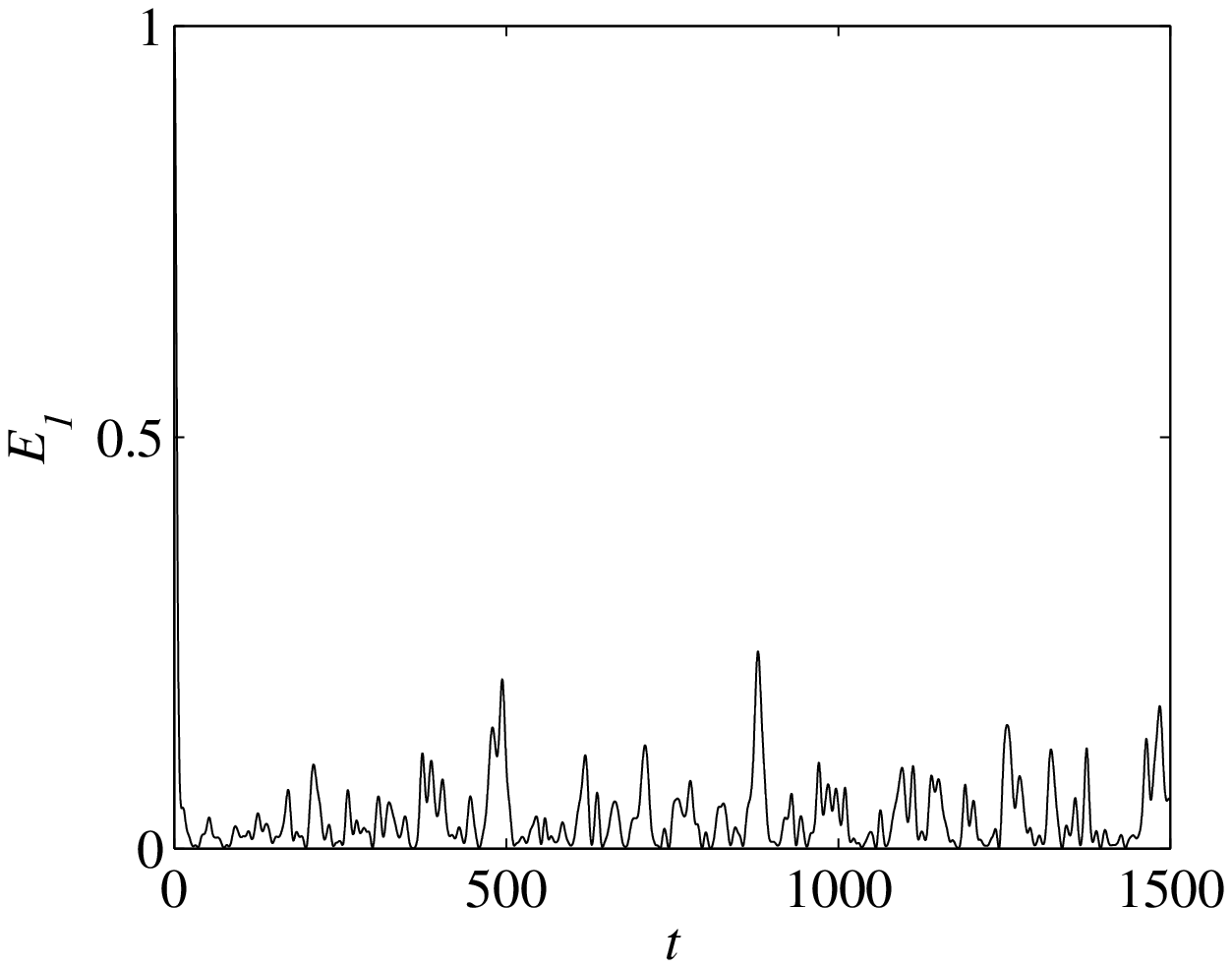} 
\includegraphics[scale=0.5]{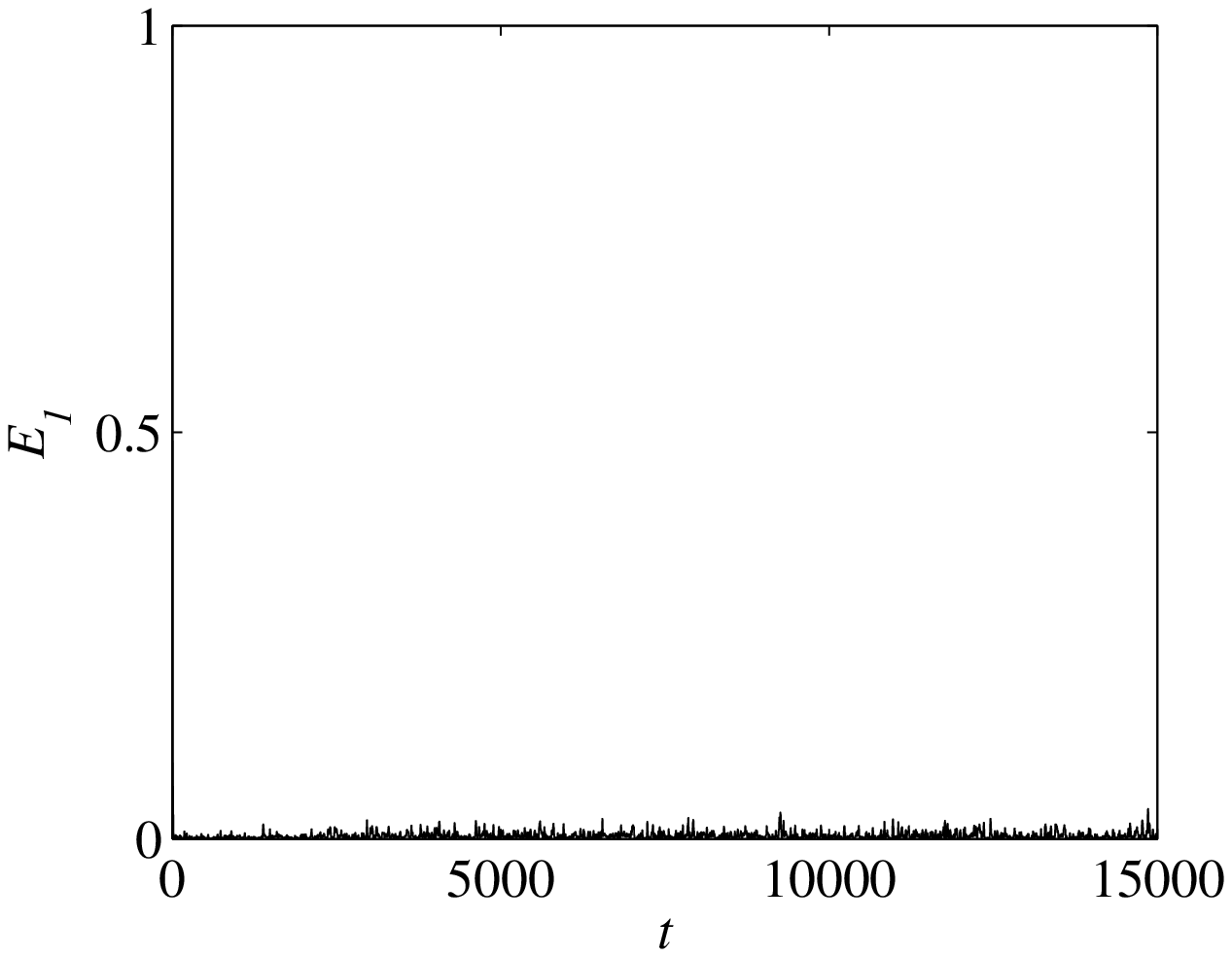} 

\textbf{(a)} \hspace{7cm} \textbf{(b)}
\vspace{-0.0cm}
\caption{$E_{1}$ as a function of time for $\hbar v_{1}=1$, $\hbar v_{k}$
randomly distributed in the interval $\left[ 0.5,1.5\right] $, $g_{k}=0.2/\sqrt{N-1}$ ($k=2$ to $N$) and
(a) $N=102$, (b) $N=1002$.}
\label{DR02}
\end{figure}

Let us start by investigating the intermediate case $g/p=0.2$. For $t=0$, we have $E_{1}=1$.
From the mathematical point of view, this result is a consequence of the
orthonormality of $\mathbf{M}$. As time progresses, the terms in the
summation in expression (\ref{Ec}) gain different phases, leading to the
decreasing of $E_{1}\left( t\right) $. As shown in the Appendix, the time
scale for this decay is given by 
\begin{equation}
\tau _{d}=\frac{1}{2\gamma \beta } \approx 3.232 ,
\end{equation}
where $\gamma $ and $\beta $ are parameters obtained through Lorentzian and
linear fittings on the points of Figs. (\ref{M02100}a) and (\ref{M02100}b), respectively.
Notice that this scale is in agreement with the evolution in Fig.
(\ref{M02100}c). A partial revival occurs around 
\begin{equation}
\tau _{r}=\frac{2\pi \left( N-1\right) }{\left( \omega _{N}-\omega
_{1}\right) }
\end{equation}
(see Fig. (\ref{M02100}d)). In order to understand this revival, we start by
observing that the distribution in Fig. (\ref{M02100}b) is almost linear.
Accordingly, the differences $\omega _{j}-\omega _{j-1}$ for $j=2$ to $N$
are, in good approximation, given by their mean value $\left( \omega
_{N}-\omega _{1}\right) /\left( N-1\right) $, and, when $t=\tau _{r}$, $%
e^{-i\omega _{j}t}$ and $e^{-i\omega _{j-1}t}$ accumulates a phase
difference close to $2\pi $; then every terms in the right hand side of Eq. (%
\ref{Ec}) will be approximately in phase. Other revivals occur for $t=n\tau
_{r}$, where $n$ is an integer greater than $1$; when $n$ increases, they go
spreading. Of course, revivals are not expected for a model of a dissipative
environment. In fact, they become unimportant when the density of
environmental modes tends to infinite, since in this limit $\tau
_{r}\longrightarrow \infty $. Another way to avoid these relatively
localized revivals is to randomly distribute the frequencies of the
environmental modes, that leads to permanent fluctuations that may be
associated to a spreading of the revival. As may be seen in Figs. (\ref{DR02}a)
and (\ref{DR02}b), these fluctuations decrease as $N$ increases, becoming
unimportant in the thermodynamic limit. 

\begin{figure}[tbp]
\vspace{-0.5cm} \hspace{0cm} 
\includegraphics[scale=0.5, angle = 0]{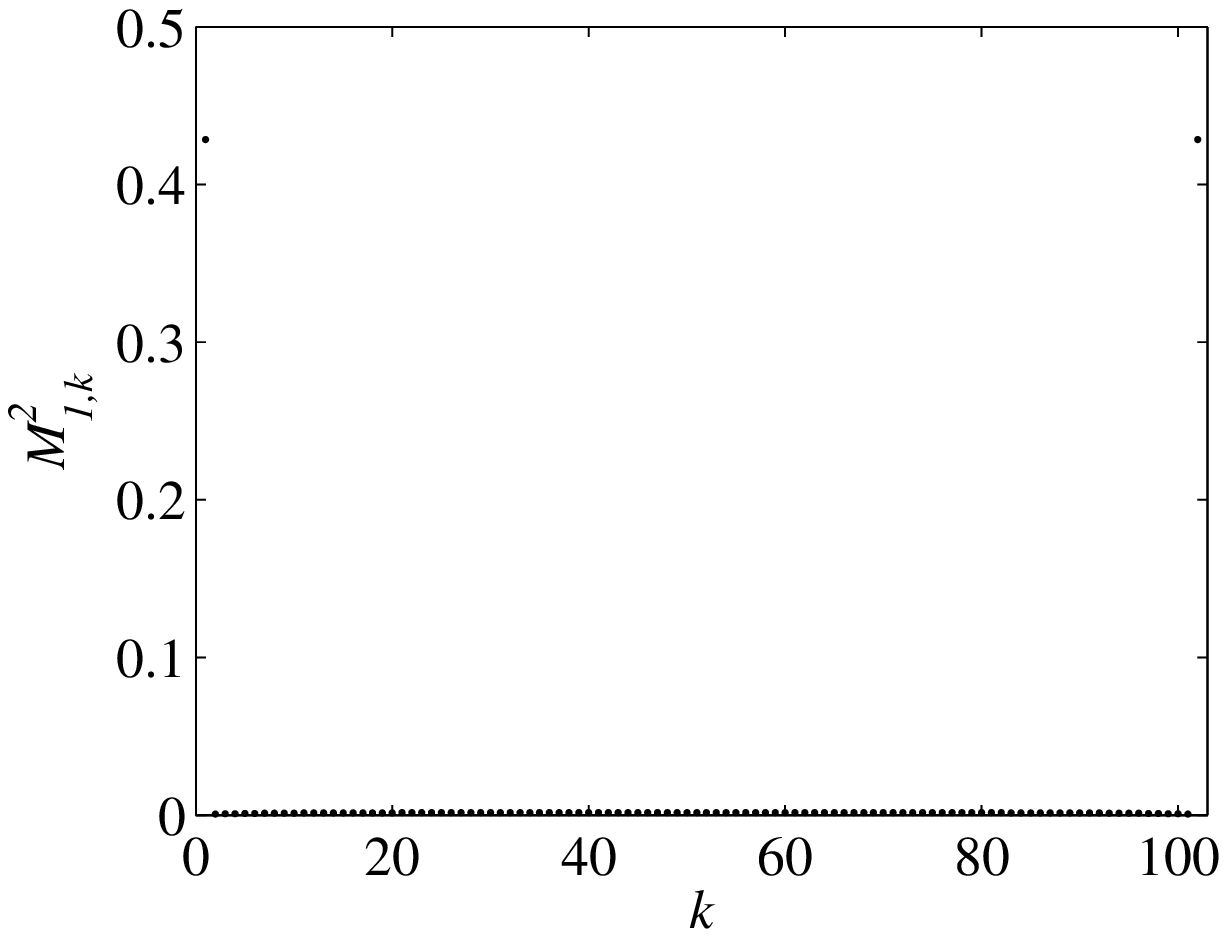} 
\includegraphics[scale=0.5, angle = 0]{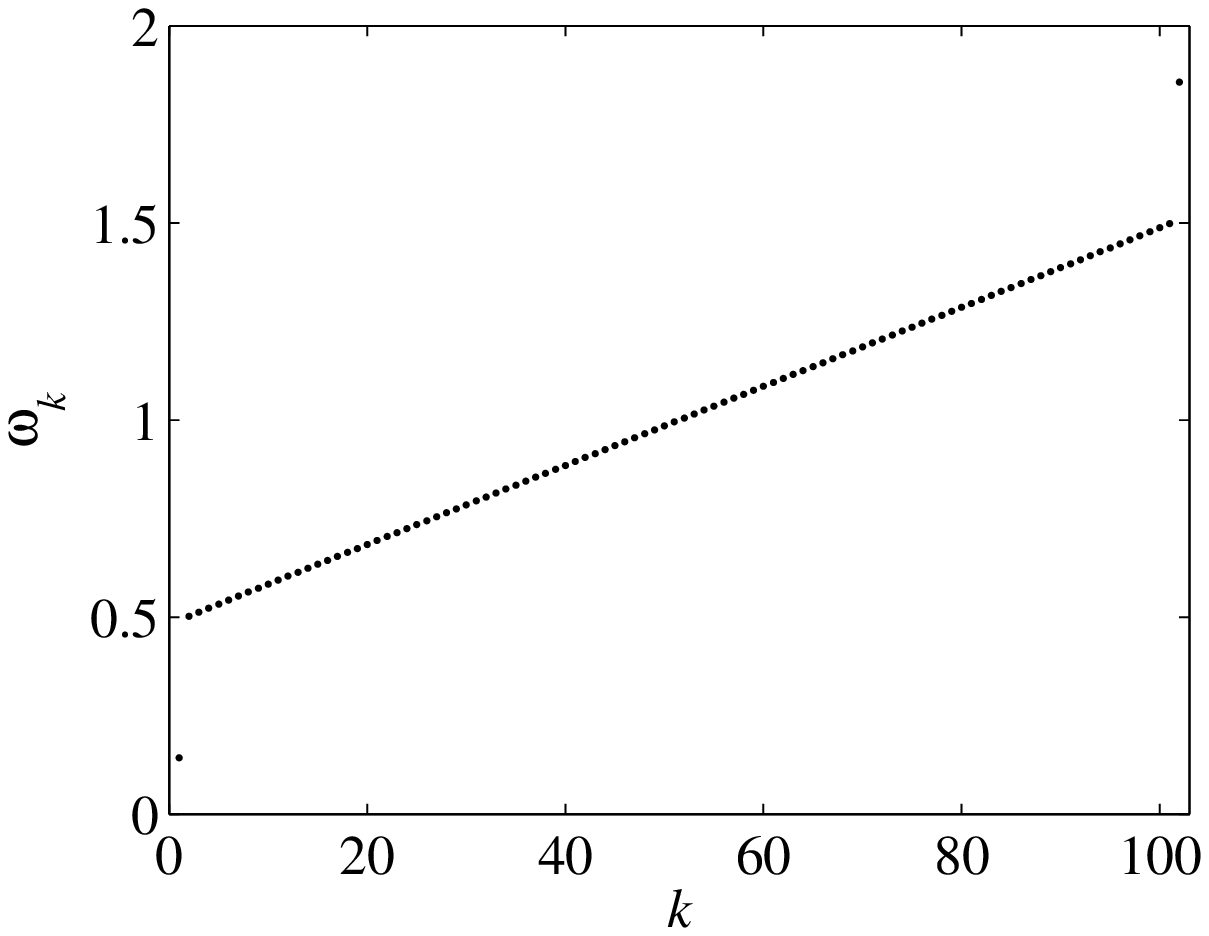}

\textbf{(a)} \hspace{7cm} \textbf{(b)}

\includegraphics[scale=0.5]{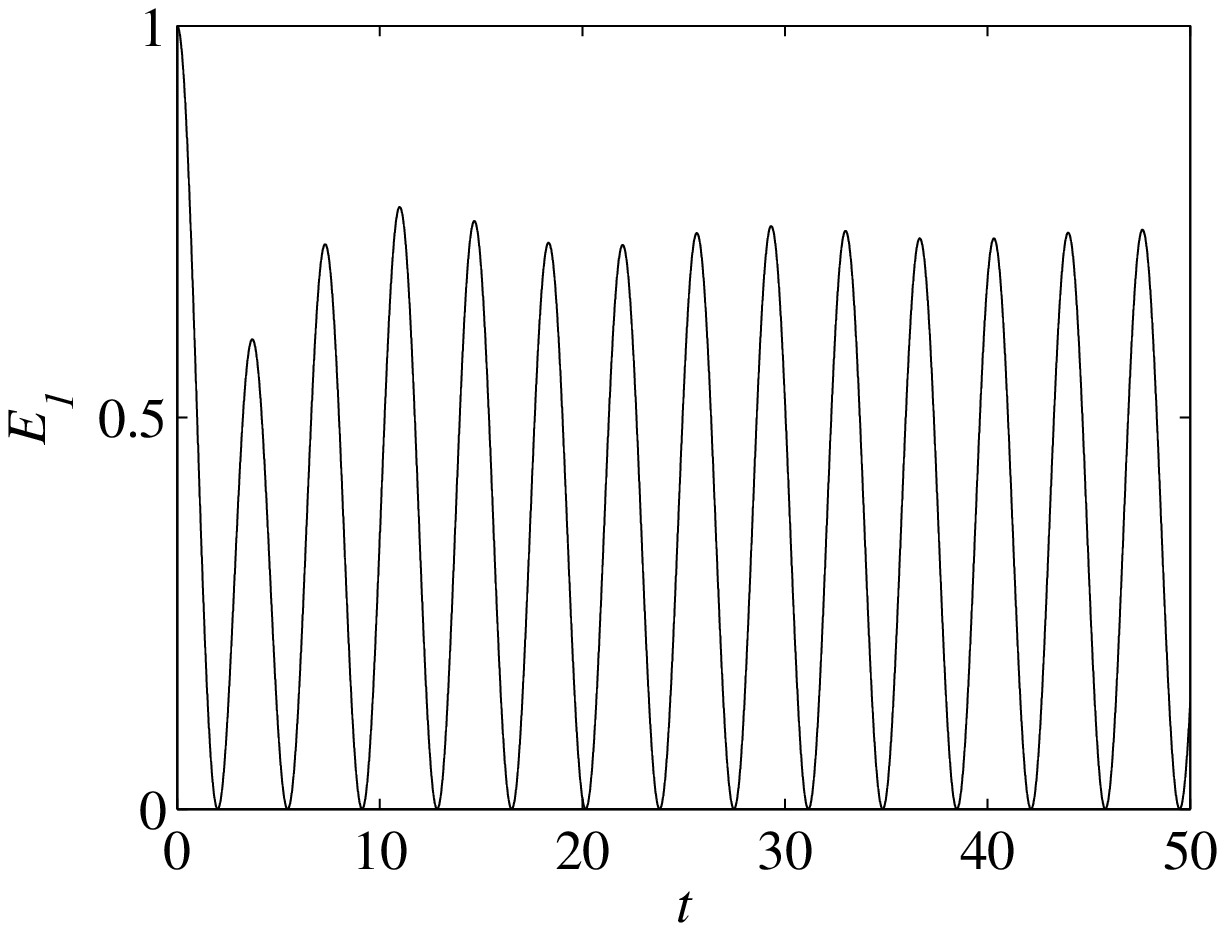}
\includegraphics[scale=0.5]{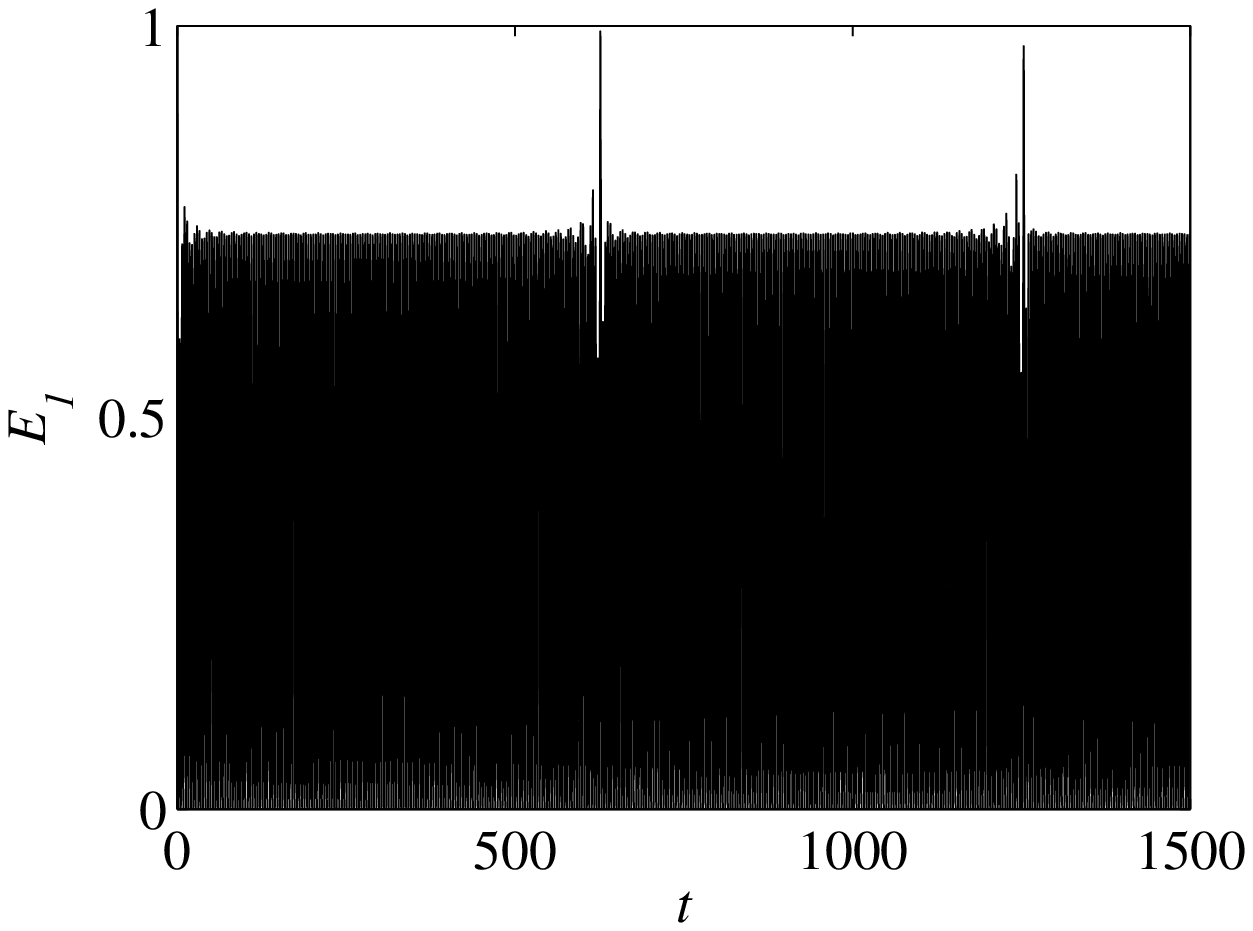}

\textbf{(c)} \hspace{7cm} \textbf{(d)}
\vspace{-0.0cm}
\caption{
(a) and (b) give the sets $\left\{ M_{1,k}^{2}\right\} $ and $\left\{ \omega _{k} \right\}$, respectively.
(c) and (d) show, in different time scales, the evolution of $E_{1}$.
For all figures, $\mathbf{H}$ is given by Eq. (\ref{Hparticular1})
with $N=102$, $p=1$ and $g=0.8$.}
\label{M08100}
\end{figure}

\begin{figure}[th]
\vspace{-0.5cm} \hspace{0cm} \includegraphics[scale=0.5]{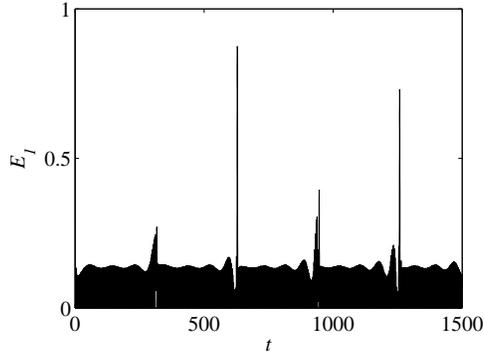} 
\vspace{-0.7cm}
\caption{$E_{1}$ as a function of time for $\mathbf{H}$ given by Eq. (\ref{Hparticular1})
with $N=102$, $p=2$ and $g=0.8$.}
\label{DP208100}
\end{figure}

In Fig. (\ref{M08100}d), we display the evolution of $E_{1}\left( t\right) $ for $%
g/p=0.8$, which may be regarded a large value for this ratio. The
oscillations of the energy between system and environment are substantial.
According to Figs. (\ref{M08100}a), (\ref{M08100}b) and (\ref{M08100}c), these are well
approximated by preserving only the most relevant terms in Eq. (\ref{Ec}): 
\begin{equation}
E_{1}\left( t\right) \approx \left| M_{1,1}^{2}e^{-i\omega
_{1}t}+M_{1,N}^{2}e^{-i\omega _{N}t}\right| ^{2}\approx 0.367\left[
1+\cos \left( 1.714t\right) \right] .
\end{equation}
Such oscillations, which are not expected for a model of dissipative
environment, are related to the finite range of environmental frequencies,
decreasing when $p$ is enhanced, as is exemplified by comparing Figs.
(\ref{M08100}d) and (\ref{DP208100}). This finitude becomes important when the coupling
strength with the environment increases.

\begin{figure}[tbp]
\vspace{-0.5cm} \hspace{0cm} 
\includegraphics[scale=0.5, angle = 0]{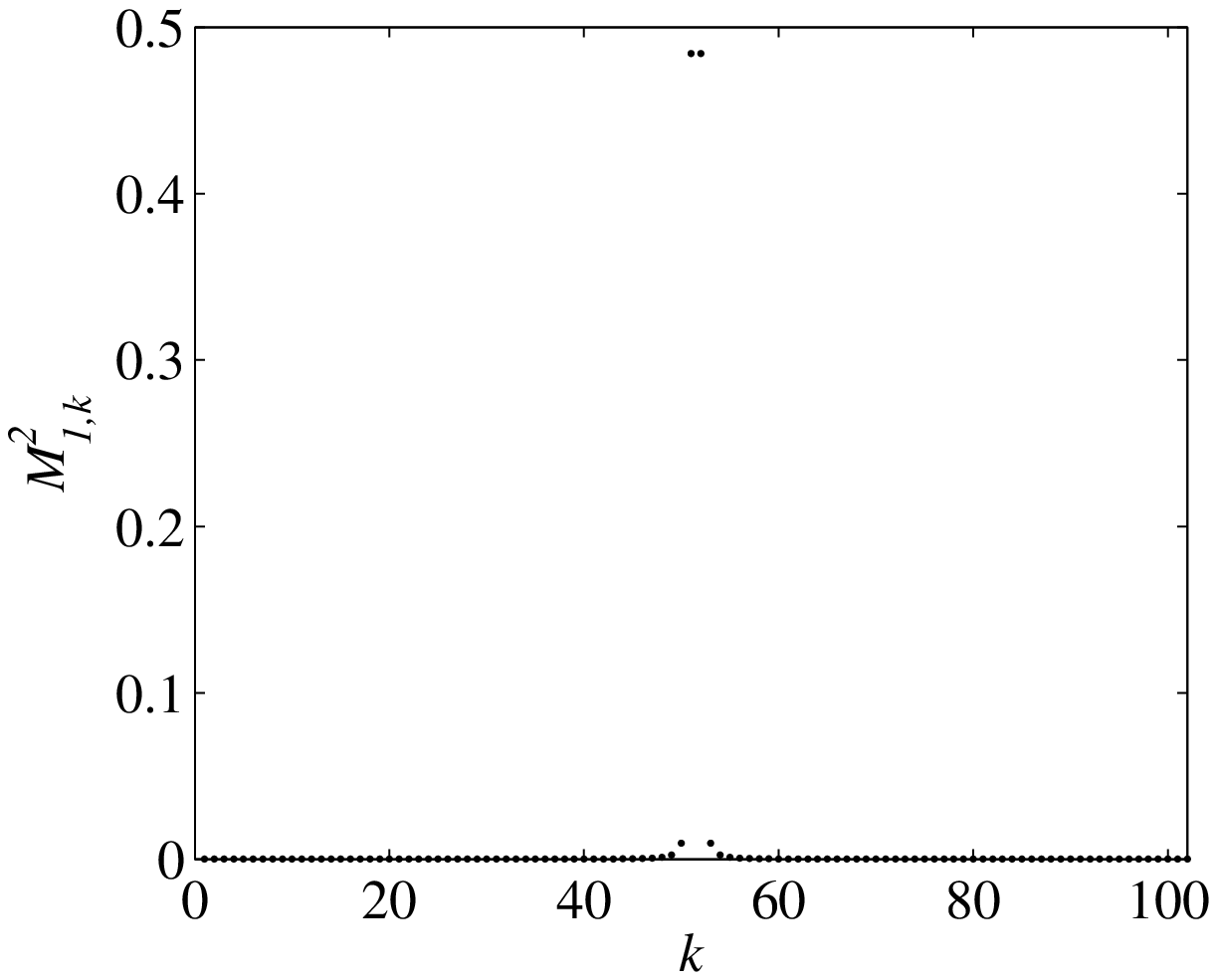} 
\includegraphics[scale=0.5, angle = 0]{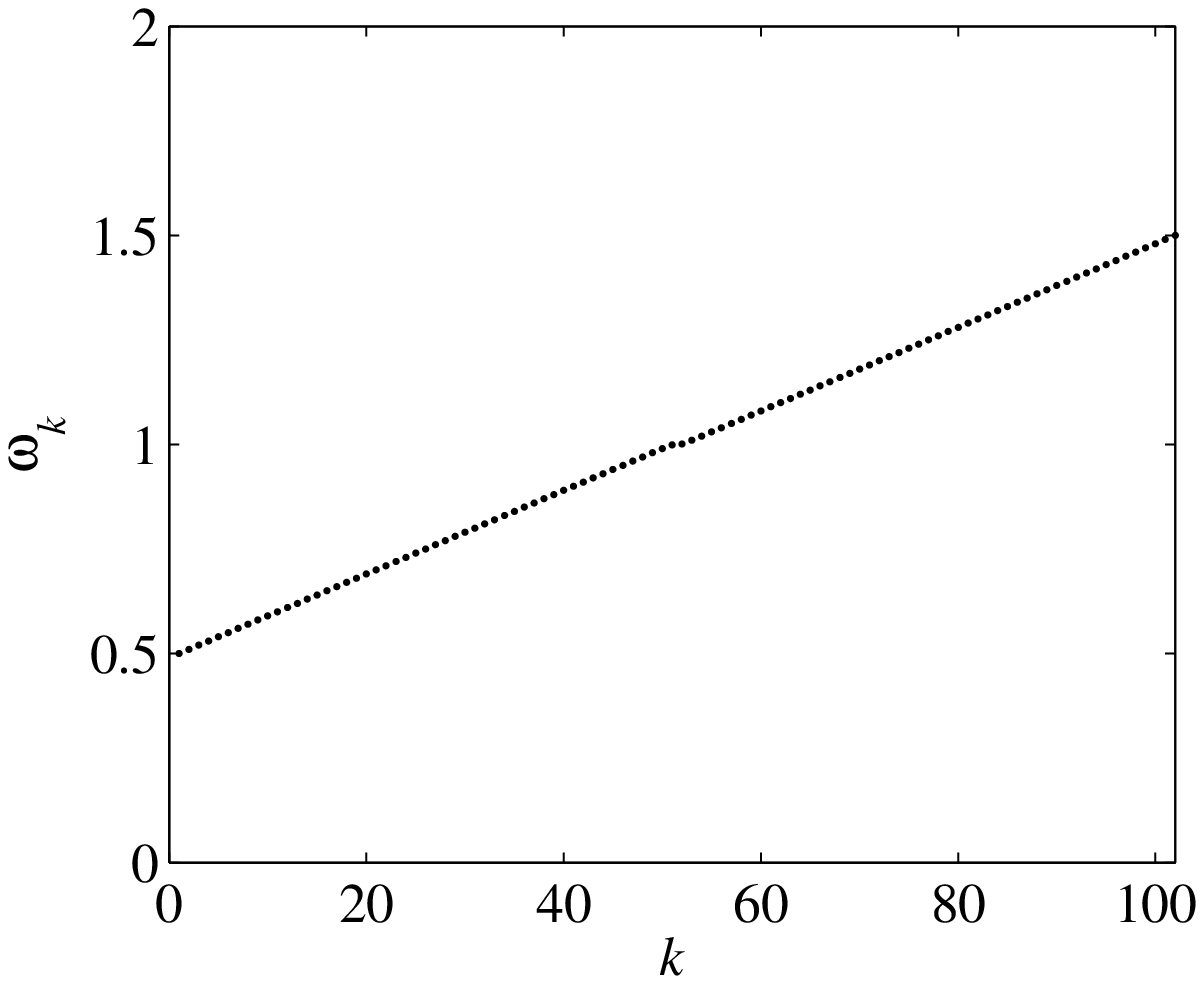}

\textbf{(a)} \hspace{7cm} \textbf{(b)}

\includegraphics[scale=0.5]{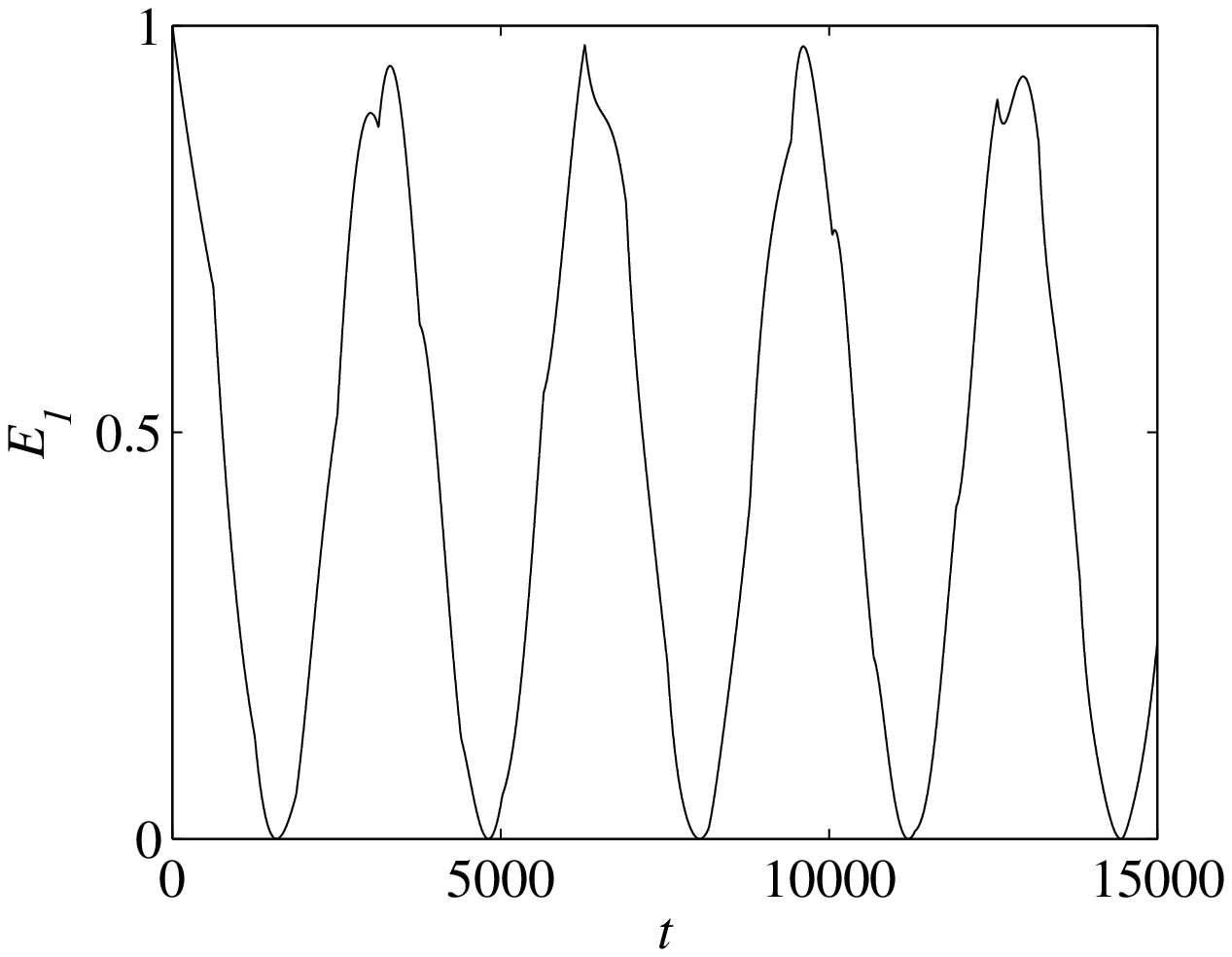}

\textbf{(c)}
\vspace{-0.0cm}
\caption{
(a) and (b) give the sets $\left\{ M_{1,k}^{2}\right\} $ and $\left\{ \omega _{k} \right\}$, respectively.
(c) shows the evolution of $E_{1}$.
For all figures, $\mathbf{H}$ is given by Eq. (\ref{Hparticular1})
with $N=102$, $p=1$ and $g=0.01$.}
\label{M001100}
\end{figure}

\begin{figure}[tbp]
\vspace{-0.5cm} \hspace{0cm} 
\includegraphics[scale=0.5, angle = 0]{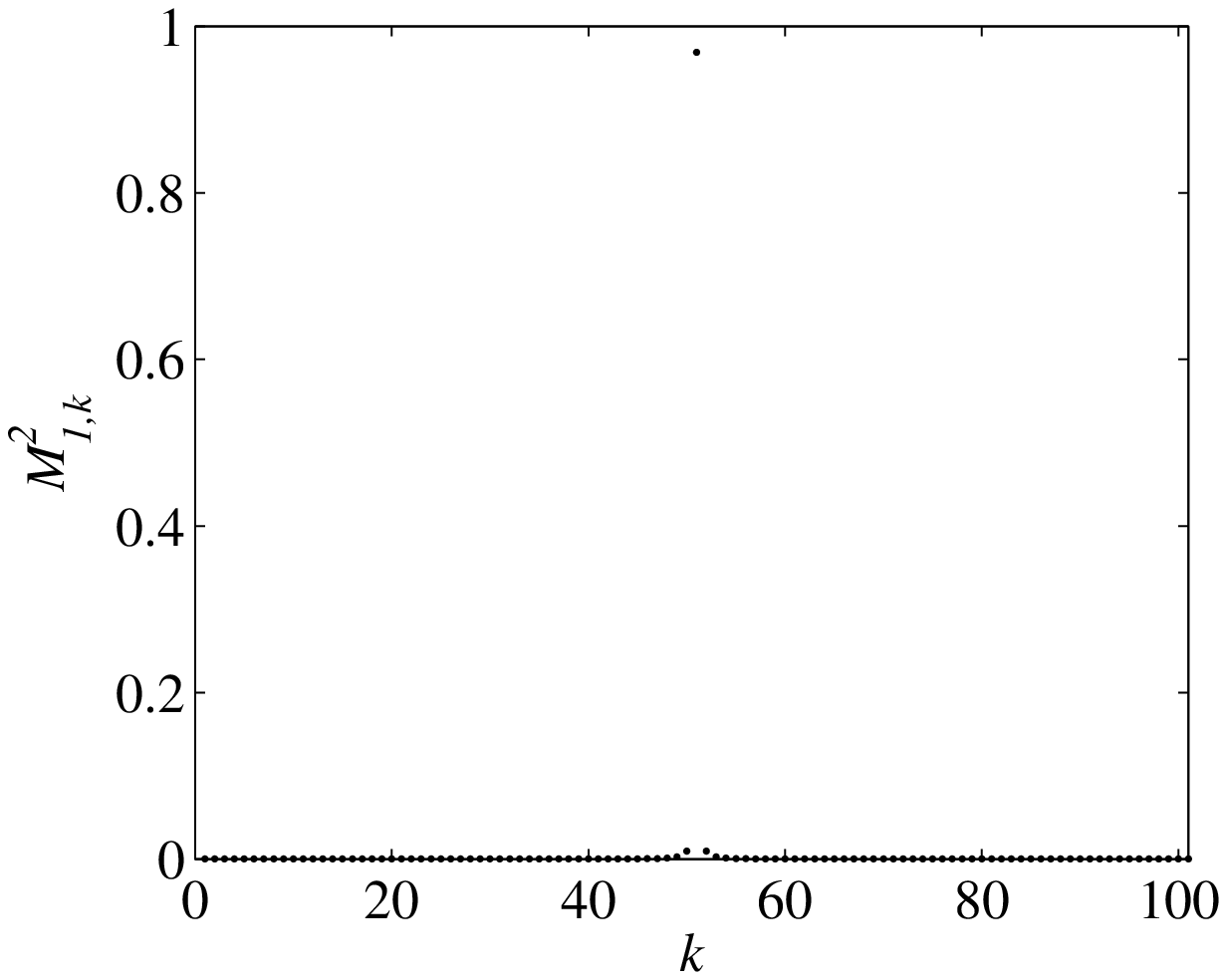} 
\includegraphics[scale=0.5, angle = 0]{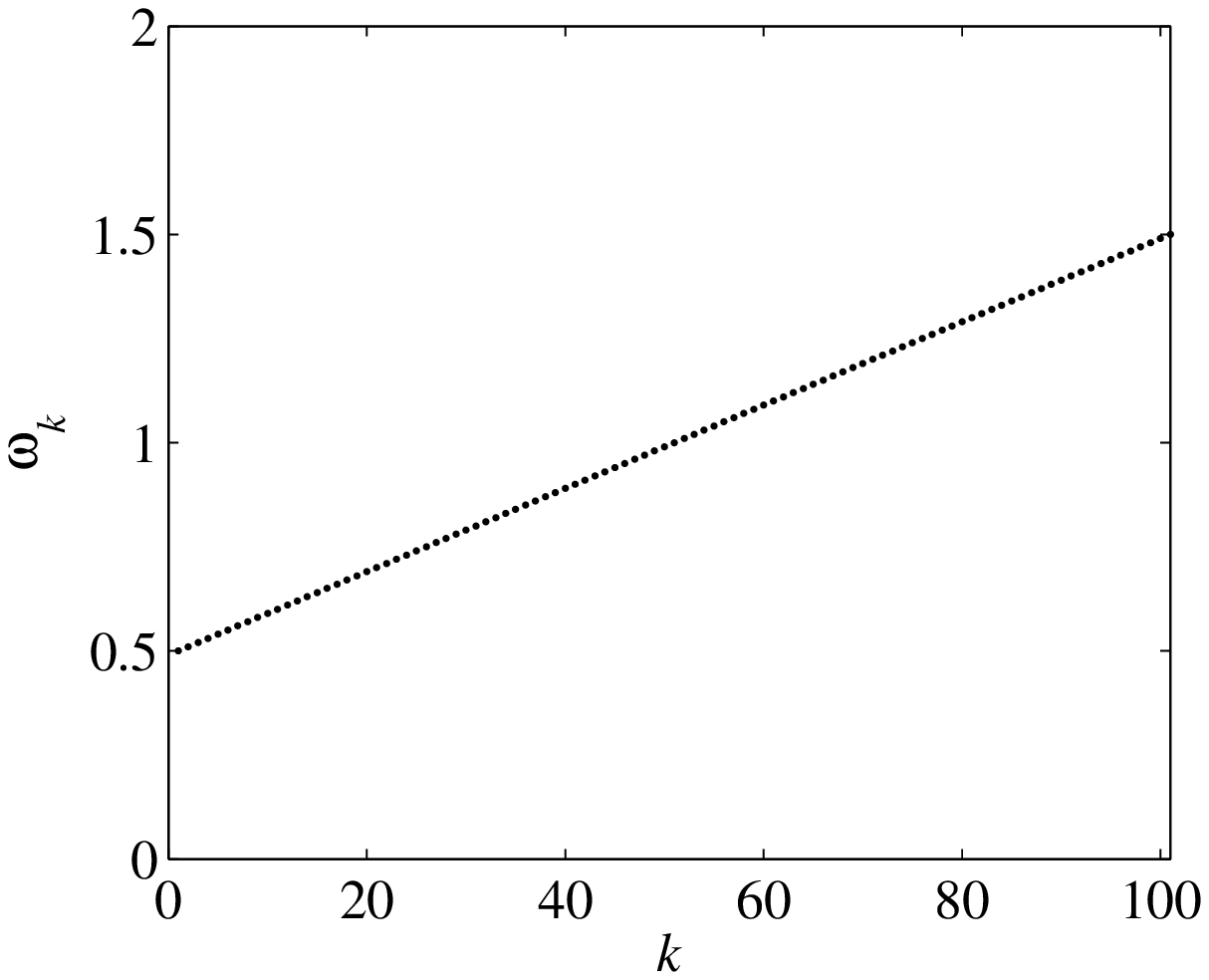}

\textbf{(a)} \hspace{7cm} \textbf{(b)}

\includegraphics[scale=0.5]{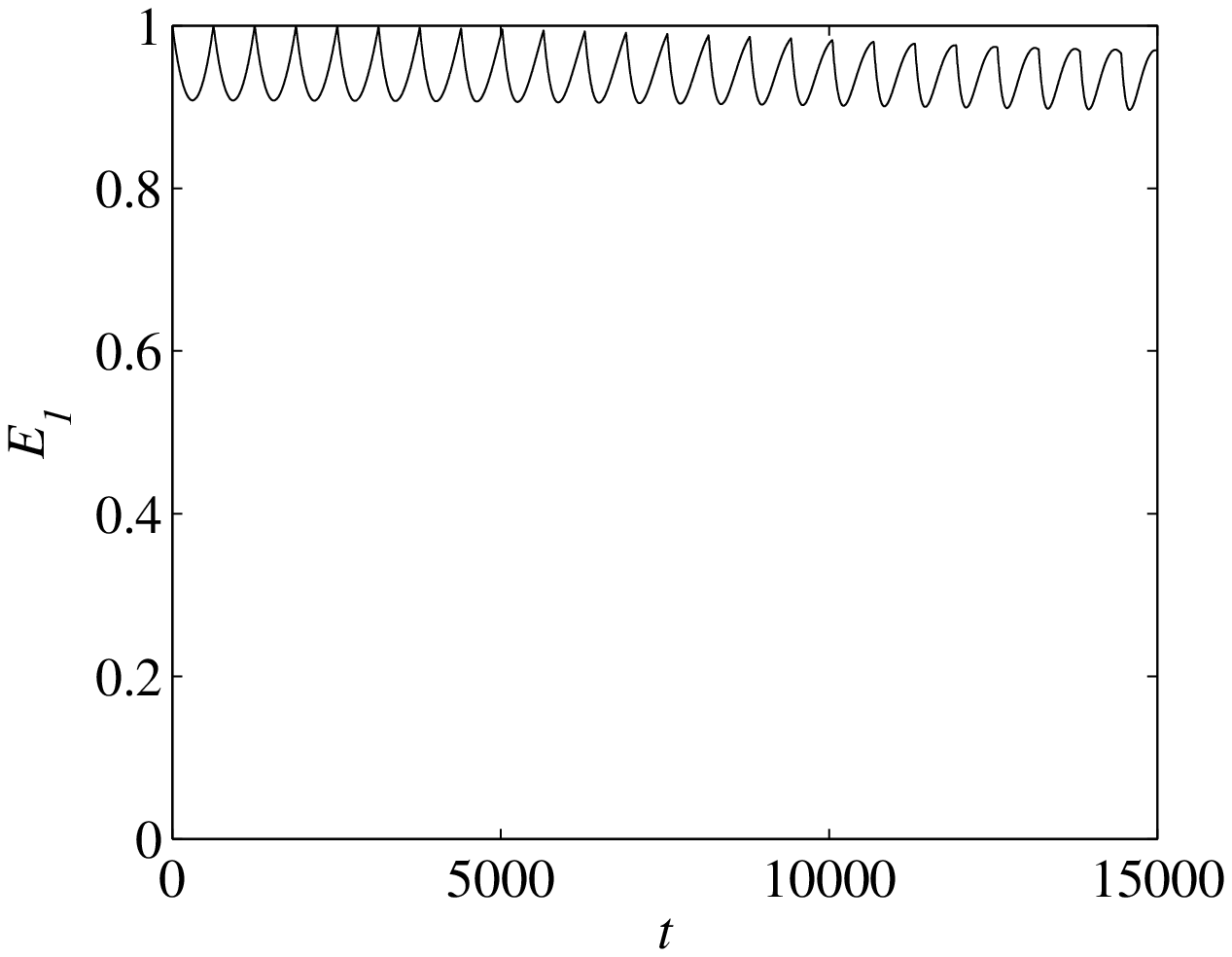}
\includegraphics[scale=0.5]{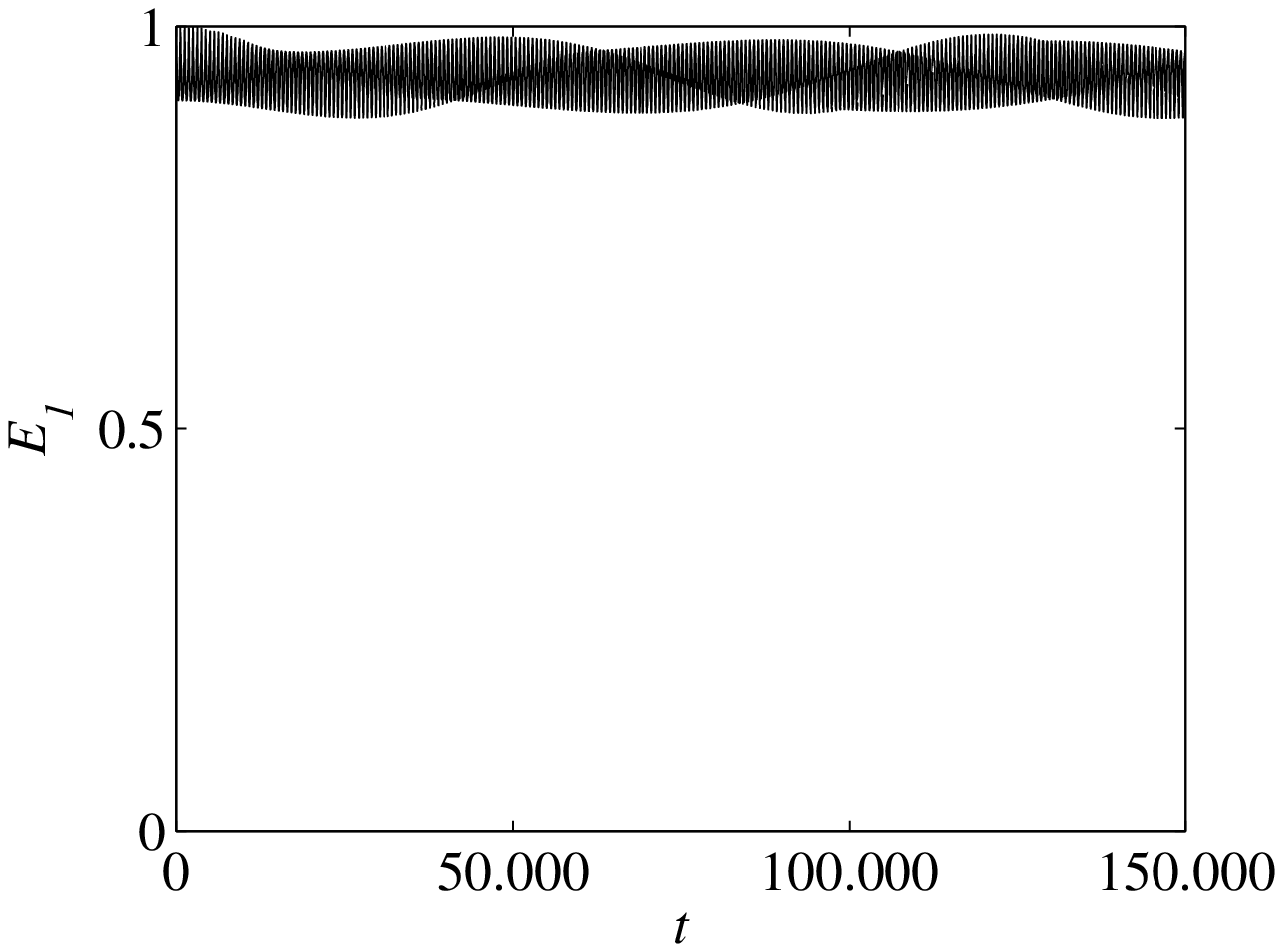}

\textbf{(c)} \hspace{7cm} \textbf{(d)}
\vspace{-0.0cm}
\caption{
(a) and (b) give the sets $\left\{ M_{1,k}^{2}\right\} $ and $\left\{ \omega _{k} \right\}$, respectively.
(c) and (d) show, in different time scales, the evolution of $E_{1}$.
For all figures, $\mathbf{H}$ has the form given in Eq. (\ref{Hparticular1}), 
except for second line and second column absent. Here, $N=101$, $p=1$ and $g=0.01$.}
\label{M00100NR}
\end{figure}

The case of weak coupling is treated in Figs. (\ref{M001100}a) and (\ref{M001100}b),
prepared with $g/p=0.01$. Now, 
\begin{equation}
E_{1}\left( t\right) \approx \left| M_{1,N/2}^{2}e^{-i\omega
_{N/2}t}+M_{1,1+N/2}^{2}e^{-i\omega _{1+N/2}t}\right| ^{2}\approx
0.469\left( 1+\cos \left[ \left( 0.002t\right) \right] \right) ,
\end{equation}
agreeing with Fig. (\ref{M001100}c). The inflection in Fig. (\ref{M001100}b)
indicates that the difference between $\omega _{N/2}$ and $\omega _{1+N/2}$
is shorter than the other $\omega _{j}-\omega _{j-1}$ differences, what
reflects the approximation to the condition $g=0$, where there is a
degenerate eigenvalue: $\omega _{N/2}=\omega _{1+N/2}=1$. The freezing of the
dynamics associated with the decrease of $g$ comes from the increasing of the period of the main oscillation. It
is noteworthy that, if the resonant environmental mode is taken out, the
dynamics for low $g/p$ will be qualitatively different (see Figs. (\ref{M00100NR}c) and (\ref{M00100NR}d)). 
As $g/p$ goes to zero, the eigenvalues $\omega _{N/2}$ and $\omega _{1+N/2}$ remain non-degenerate
and $M_{1,\left(N+1\right) /2}^{2}$ approaches the unit.
With the help of Figs. (\ref{M00100NR}a) and (\ref{M00100NR}b), we see that 
\begin{eqnarray}
E_{1}\left( t\right) &\approx& \left| 
M_{1,\left(N+1\right) /2-1}^{2}e^{-i\omega_{\left(N+1\right)/2-1}t}+
M_{1,\left(N+1\right) /2}^{2}e^{-i\omega_{\left(N+1\right)/2}t}+
M_{1,\left(N+1\right) /2+1}^{2}e^{-i\omega_{\left(N+1\right)/2+1}t}
\right| ^{2} \notag \\
&\approx&
\left( 0.968 + 0.019 \cos \left( 0.010t\right) \right)^{2} ,
\end{eqnarray}
where we used $M_{1,\left(N+1\right) /2-1}^{2} \approx M_{1,\left(N+1\right) /2+1}^{2}$ and
$\omega_{\left(N+1\right)/2}-\omega_{\left(N+1\right)/2-1} \approx \omega_{\left(N+1\right)/2+1}-\omega_{\left(N+1\right)/2}$.
The standstill of $E_{1}\left( t\right) $ is achieved through the increasing of 
$M_{1,\left(N+1\right) /2}^{2}$.

\section{Conclusions}

Irreversibility is often expected for the dynamics of a system affected by the environment. This is usually achieved by considering the set of environmental degrees of freedom in certain limits, as dense spectrum spread in an infinite interval. In order to investigate the way the system's dynamics approaches the irreversibility, the
environments we analyzed here are finite. As we show, the expected behaviors may be reached through different ways. 
The dynamics we calculated depends crucially on the sets of eigenvalues and eigenvectors of the Hamiltonian of system plus environment. Although these sets are not obtainable analytically, we analyzed their structures by considering them as varying between two analytically tractable cases. This way, the numerically calculated dynamics can be more deeply understood.

Since the bath discussed is at null temperature, we expect that it produces dissipation. 
In order to obtain such a dynamics, the relation between the interval where the environment frequencies are distributed and the strength of the coupling to the environment must be properly chosen. 
If the coupling dominates, the environment approaches the situation where all oscillators have the same frequency, leading to a collective behavior of the environment oscillators that induces an evolution similar to the one concerning a single oscillator highly coupled to the system, i.e., fast periodic energy flow between system and environment.
In an suited situation, the relation between coupling strength and bandwidth must be such that the effects of the oscillators out of this bandwidth are negligible, due to detuning. Approaching this case, the fast oscillations described above progressively decrease, until we get a dynamics that resembles dissipation. The role of a large (in the limit, infinite) bandwidth in the model is to avoid the collective behavior of the oscillators of the environment.

Since the bath is finite, the energy that flows from the system returns, in part, in finite times. The form of this flow back depends on the distribution of the bath modes: for equally spaced frequencies, there are well defined partial revivals; for randomly distributed frequencies, there are no revivals, but the residual oscillations are relatively large. As we approach the condition of dense spectrum, irreversibility is achieved, for equally spaced frequencies, through the growing of the revival time, while, for randomly distributed frequencies, through the decrease of the residual oscillations. 

For relatively strong coupling, the importance of each single bath oscillator for the system dynamics is small. Nevertheless, for weak coupling, the presence of one single resonant mode is crucial for the definition of the shape of the dynamics. The portion of energy that two coupled linear oscillators can exchange depends on the relation between coupling and detuning. For weak coupling, this portion will be small for non-resonant oscillators, while the whole energy is aways exchangeable for the resonant one. Accordingly, if there is a resonant mode, the interaction of the system with this mode dominates. Thus, the dynamics approaches the one where the system interacts with a single loosely coupled oscillator, leading to a large period for the flow of energy. This period increases with the number of oscillators, paving the way for irreversibility. Already for absent resonant mode, only a small part of the energy can flow between the system and the environment. Exact resonance is hardly expected in a real case, but if there is one oscillator for which the ratio between dephasing and coupling is low and much lower than for the other, the dynamics may be close to the one with the resonant mode.

The limit conditions leading to irreversibility can be reached by modeling the bath with different frequency distributions. Although the dynamics induced are the same when the limit conditions are assumed, if we move away from them the evolution of the system can vary qualitatively.

\acknowledgements{The authors acknowledge the Brazilian agencies CNPq and Fapemig
for partial financial support.}

\appendix
\section{}

According to Eq. (\ref{Ec}), the energy of oscillator $O_{1}$ may be written
as 
\begin{equation}
E_{1}\left( t\right) =\hbar v_{1}\left| \sigma \left( t\right) \right| ^{2},
\label{E1}
\end{equation}
where 
\begin{equation}
\sigma \left( t\right) =\overset{N}{\underset{j=1}{\sum }}%
M_{1,j}^{2}e^{-i\omega _{j}t}\text{.}  \label{sigma}
\end{equation}
In this Appendix, we employ the concept of Fourier transform of a function $%
f\left( t\right) $, 
\begin{equation}
F\left( k\right) =\frac{1}{\sqrt{2\pi }}\overset{+\infty }{\underset{-\infty 
}{\int }}f\left( t\right) e^{ikt}dt,  \label{TF}
\end{equation}
and its inverse, 
\begin{equation}
f\left( t\right) =\frac{1}{\sqrt{2\pi }}\overset{+\infty }{\underset{-\infty 
}{\int }}F\left( k\right) e^{-ikt}dk,  \label{TIF}
\end{equation}
to analyze $\sigma \left( t\right) $. We will explore possible relations
between eqs. (\ref{sigma}) and (\ref{TIF}).

\begin{figure}[tbp]
\vspace{-0.5cm} \hspace{0cm} 
\includegraphics[scale=0.5, angle = 0]{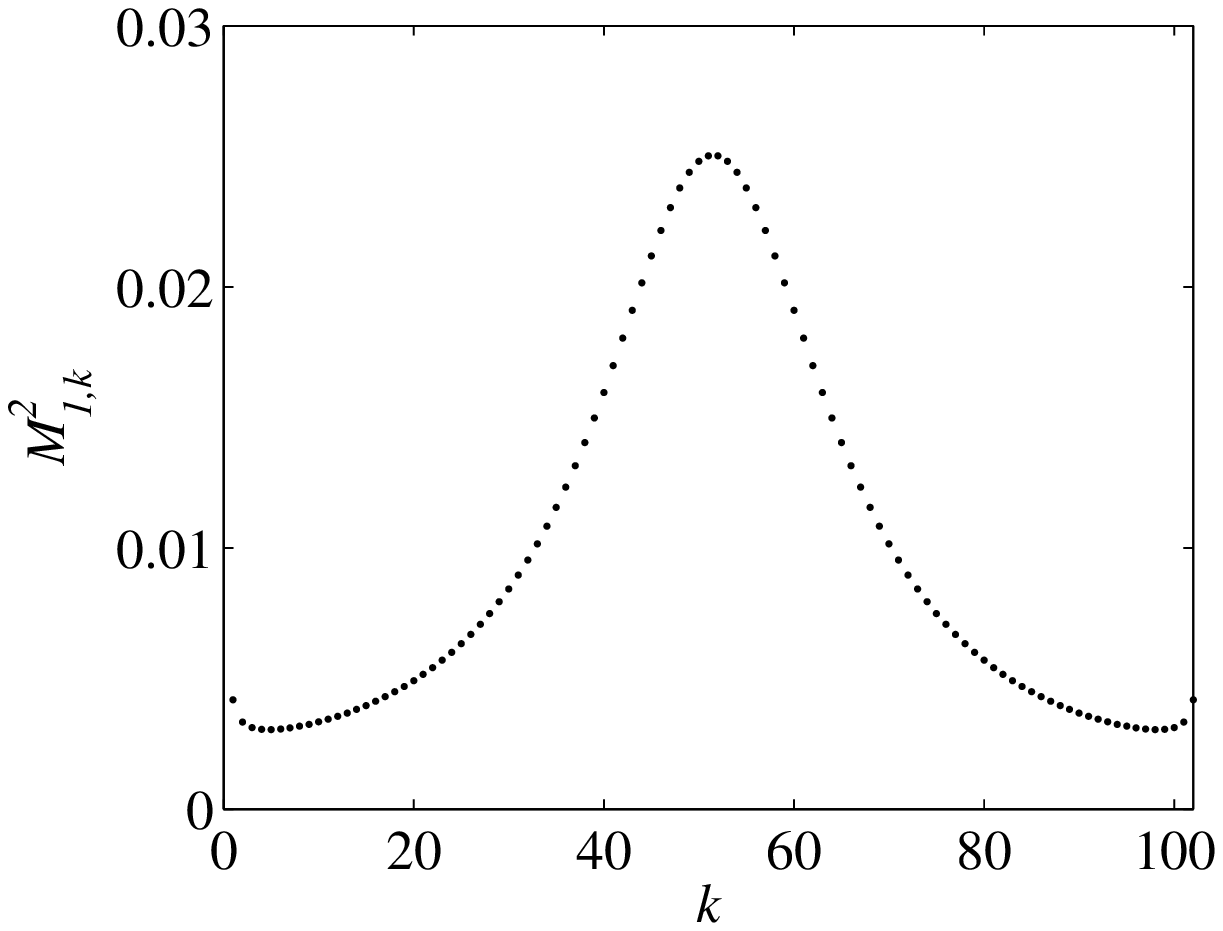} 
\includegraphics[scale=0.5, angle = 0]{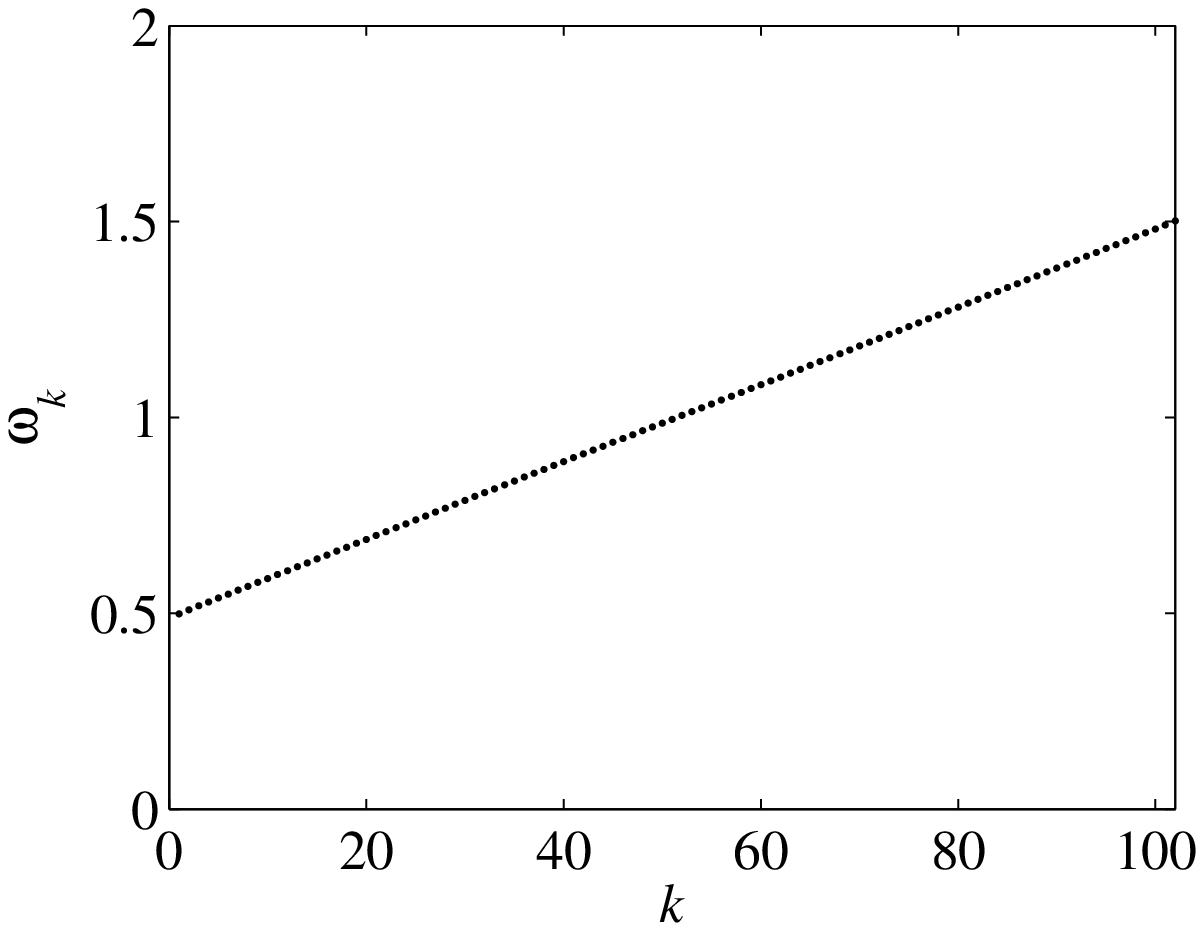}

\textbf{(a)} \hspace{7cm} \textbf{(b)}

\includegraphics[scale=0.5]{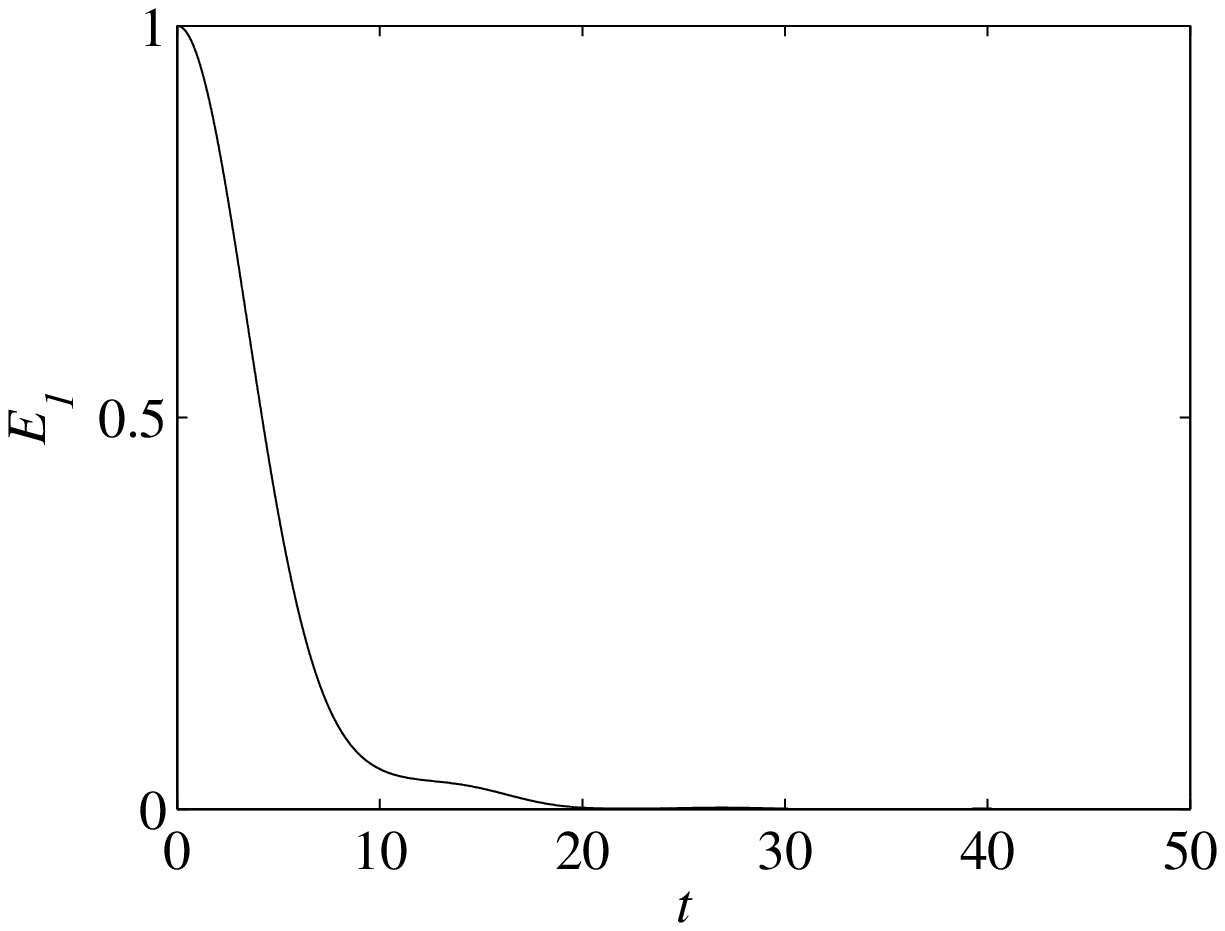}
\includegraphics[scale=0.5]{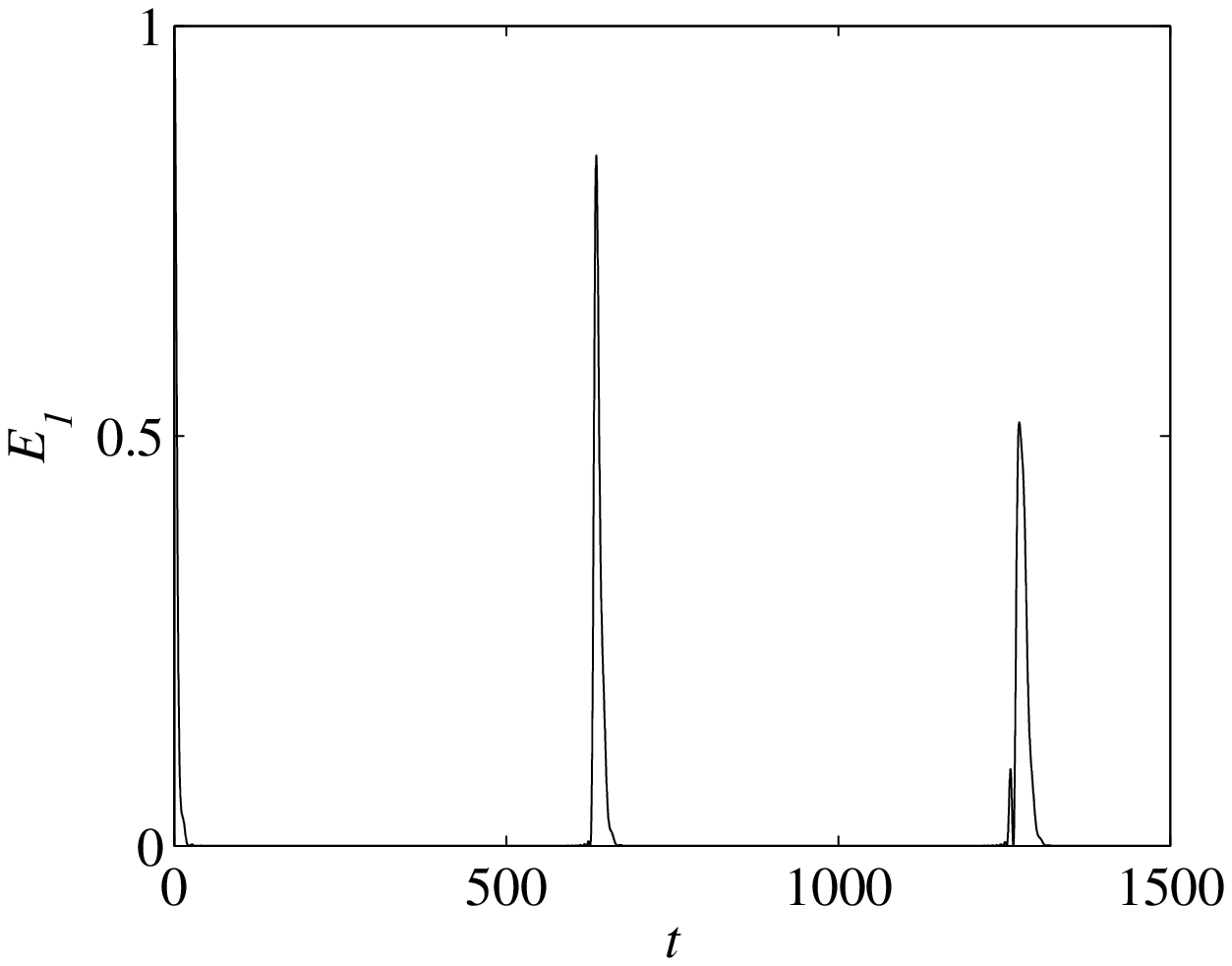}

\textbf{(c)} \hspace{7cm} \textbf{(d)}
\vspace{-0.0cm}
\caption{
(a) and (b) give the sets $\left\{ M_{1,k}^{2}\right\} $ and $\left\{ \omega _{k} \right\}$, respectively.
(c) and (d) show, in different time scales, the evolution of $E_{1}$.
For all figures, $\mathbf{H}$ is given by Eq. (\ref{Hparticular1})
with $N=102$, $p=1$ and $g=0.2$.}
\label{M02100}
\end{figure}

For the parameters of Fig. (\ref{M02100}), it may be assumed that $\omega
_{j}\approx \omega _{0}+\beta j$, where $\omega _{0} \approx 0.4902 $ 
and $\beta \approx 9.899 \times 10^{-3}$ 
are obtained through linear regression. Using this approximation, we can
write 
\begin{equation}
\sigma \left( t\right) \approx e^{-i\omega _{0}t}\overset{N}{\underset{j=1}{%
\sum }}M_{1}^{2}\left( j\right) e^{-i\beta jt},
\end{equation}
where $M_{1}\left( x\right) $ is a continuous function defined for every
real number $x$ such that $M_{1}\left( j\right) =M_{1,j}$ for $j=1,2,\cdots
,N$. By changing the summation into an integral and considering that $%
M_{1}^{2}\left( x\right) $ is negligible for $x<0$ or $x>N$, we get 
\begin{equation}
\sigma \left( t\right) \approx e^{-i\omega _{0}t}\overset{+\infty }{%
\underset{-\infty }{\int }}M_{1}^{2}\left( x\right) e^{-it\beta x}dx,
\end{equation}
which, after the variables transformation $k=\beta x$, leads to 
\begin{equation}
e^{i\omega _{0}t}\sigma \left( t\right) \approx \frac{1}{\sqrt{2\pi }}%
\overset{+\infty }{\underset{-\infty }{\int }}\frac{\sqrt{2\pi }}{\beta }%
M_{1}^{2}\left( \frac{k}{\beta }\right) e^{-ikt}dk.  \label{phi}
\end{equation}
By comparing eqs. (\ref{TIF}) and (\ref{phi}), we see that 
\begin{equation}
\Phi \left( k\right) =\frac{\sqrt{2\pi }}{\beta }M_{1}^{2}\left( \frac{k}{%
\beta }\right) 
\end{equation}
is nearly the Fourier transform of 
\begin{equation}
\phi \left( t\right) =e^{i\omega _{0}t}\sigma \left( t\right) .
\end{equation}
Of course, we can substitute $\sigma \left( t\right) $ for $\phi \left(
t\right) $ in Eq. (\ref{E1}) without any modification in $E_{1}\left(
t\right) $. 

Now we assume that $M_{1}^{2}\left( x\right) $ is approximately the
Lorentzian function 
\begin{equation}
M_{1}^{2}\left( x\right) \approx \frac{A\gamma }{\left( x-x_{0}\right)
^{2}+\gamma ^{2}},
\end{equation}
where $A \approx 0.3886 $, $\gamma \approx 15.63 $ and $x_{0} \approx 51.50 $ are obtained by fitting
the data in Fig. (\ref{M02100}). Thus, 
\begin{equation}
\Phi \left( k\right) \approx \frac{A\gamma \beta \sqrt{2\pi }}{\left(
k-\beta x_{0}\right) ^{2}+\left( \beta \gamma \right) ^{2}},
\end{equation}
what is the Fourier transform of  
\begin{equation}
\phi \left( t\right) \approx A\pi e^{ix_{0}t}e^{-\gamma \beta \left|
t\right| },
\end{equation}
leading to 
\begin{equation}
E_{1}\left( t\right) \approx \hbar \pi ^{2}v_{1}A^{2}e^{-2\gamma \beta
\left| t\right| }.
\end{equation}

\newpage

\vspace{-0.0cm}

\vspace{-0.0cm}

\vspace{-0.0cm}

\vspace{-0.0cm}

\vspace{-0.0cm} 

\vspace{-0.0cm}

\vspace{-0.0cm}
\end{document}